\documentclass[pra,twocolumn,nofootinbib,eqsecnum,floatfix,showpacs]{revtex4}
\usepackage{bm} 
\usepackage{amsmath}

\begin{document}

\title{{\bf Cosmological Measures without Volume Weighting}
\thanks{Alberta-Thy-16-08, arXiv:0808.0351 [hep-th]}}

\author{Don N. Page}
\email{don@phys.ualberta.ca}

\affiliation{Theoretical Physics Institute\\
Department of Physics, University of Alberta\\
Room 238 CEB, 11322 -- 89 Avenue\\
Edmonton, Alberta, Canada T6G 2G7}

\date{2008 September 16}

\begin{abstract}

Many cosmologists (myself included) have advocated volume weighting for the
cosmological measure problem, weighting spatial hypersurfaces by their volume. 
However, this often leads to the Boltzmann brain problem, that almost all
observations would be by momentary Boltzmann brains that arise very briefly as
quantum fluctuations in the late universe when it has expanded to a huge size,
so that our observations (too ordered for Boltzmann brains) would be highly
atypical and unlikely.  Here it is suggested that volume weighting may be a
mistake.  Volume averaging is advocated as an alternative.  One consequence
may be a loss of the argument that eternal inflation gives a nonzero
probability that our universe now has infinite volume.

\end{abstract}


\maketitle

\section{Introduction}

One of the most serious problems of theoretical cosmology today is the {\it
measure problem}, the problem of how to make statistical predictions for
observations in a universe that may be so large that almost all theoretically
possible observations actually occur somewhere.  One would like to be able to
calculate what fraction each possible class of observations makes out of all
possible observations.  However, this is problematic if the universe is
infinitely large and if there are infinitely many observations of each class. 
Then one has the ambiguities of taking the ratios of infinite quantities.

This ambiguity of dividing infinity by infinity occurs not only for open
universes that automatically have infinite volume (and hence presumably
infinitely many observations, assuming a nonzero average density of observers),
but also for closed universes that have eternal inflation to make them
arbitrarily large.  A huge effort
[1-47]
has gone into proposing procedures for making well-defined ratios of the
resulting infinities of different classes of observations.

One particular challenge has come from a consideration of Boltzmann brains
[48-74],
which are putative observers that can apparently form from thermal and/or vacuum
fluctuations arbitrarily late in a universe that lasts forever.  If the
spacetime (perhaps after being regularized to just a finite comoving volume if
the actual total spatial volume is infinite) has only a finite four-volume where
ordinary observers like us can live (e.g., during the finite lifetime of stars),
but if it lasts infinitely long and if Boltzmann brains can form at a nonzero
(even if extremely small) rate per four-volume, then it seems there will be an
infinite number of Boltzmann brains in comparison with us ordinary observers. 
Making the plausible assumption that only a very tiny fraction of Boltzmann
brain observations are similar to our ordered observations (so that we can
statistically exclude the hypothesis that we are Boltzmann brains), then since
Boltzmann brains would infinitely dominate over ordinary observers, observations
like ours would form only a very tiny fraction of the whole and so would be
extremely unlikely.

Thus Boltzmann brains appear to be a {\it reductio ad absurdum} for present
cosmological theories that allow ordinary observers in only an infinitesimally
tiny fraction of spacetime and that allow infinitely more Boltzmann brains to
form from fluctuations throughout a spacetime that lasts forever.  It is not
that cosmologists have lost their brains in considering Boltzmann brains
\cite{NYTimes}, since few of us believe that Boltzmann brains really infinitely
dominate over ordinary observers.  Instead, we realize that Boltzmann brains
form a paradox rather analogous to the ultraviolet catastrophe that plagued the
classical thermodynamics of radiation before Planck's introduction of the
quantum that cured that problem.  Now we need a suitable solution to the measure
problem to cure the putative catastrophe of Boltzmann brains.

Since Boltzmann brains are expected to form at only extremely tiny rates, such
as perhaps $\sim e^{-10^{42}}$ from vacuum fluctuations of a 1 kg brain for a
time equal to the light travel time across it \cite{Page06b}, one way to avoid
the problem is for the universe to have a finite lifetime
\cite{Page06a,BF,Page06c,Page06d,ADSV}.  With a putative decaying universe,
the problem might be solved (so long as what the universe decays into does not
itself have too many Boltzmann brains).  However, this puts severe restrictions
on the decay rate.

The tightest restriction would come if the universe asymptotically expands
exponentially, as would be the prediction from the currently observed cosmic
acceleration if the dark energy driving the expansion has constant energy
density (e.g., a cosmological constant, or the bottom of a potential well for a
scalar field).  If one imagined that such an asymptotically de Sitter universe
decays deterministically at a fixed time, it would need to decay within about
$10^{52}$ years, before the total number of Boltzmann brains per comoving volume
would dominate over ordinary observers.  However, if the universe decayed
quantum mechanically with an uncertain time of the actual decay, one would need
the expected decay time to be less than about 20 billion years
\cite{Page06a,Page06d,ADSV} in order that the expectation value of the
four-volume per comoving volume not be infinite and lead to an infinite
domination by Boltzmann brains.  Although such an astronomically rapid decay is
not directly ruled out observationally, it would apparently require unnaturally
fine tuning \cite{Page06a,Page06d,ADSV}.

There are by now quite a number of other proposed solutions to the Boltzmann
brain paradox
[48-74],
but all of them, including my own suggestions, seem to
me rather unnatural.  The main problem seems to arise from the rather plausible
deduction that there should be an infinite number of Boltzmann brains that arise
in any spacetime that lasts forever, especially if it also expands by an
unbounded amount.

Here I propose that at least a major part of the Boltzmann brain problem could
be averted if one abandoned the volume weighting that seems to imply that there
is more weight for Boltzmann brain observations for spaces of larger volume.  I
suggest replacing volume weighting by the assumption that one should average
over the volume of space, so that two spatial hypersurfaces with the same
density of each class of observations, but with different spatial volumes,
would give the same weights for these classes of observations (if the two
hypersurfaces have the same quantum amplitudes).

One would still need to sum or average over all spatial hypersurfaces in the
quantum state the result that one gets for the volume-averaged weights for each
class of observation on each spatial hypersurface.  It is less clear to me what
is the most natural way to do this sum or average over different hypersurfaces,
so that there is still a potential Boltzmann brain problem from a sum or
integral over an infinite number of different hypersurfaces, such as in a
universe that lasts forever.  This problem might be averted if the universe does
have a finite lifetime \cite{CDGKL,FL}, which with volume averaging rather than
volume weighting might be very long and not requiring fine tuning, though there
do remain potential problems with this that I shall discuss below.  Another idea
that I shall pursue below is in a classical approximation to restrict to closed
spatial hypersurfaces of constant trace of the extrinsic curvature (i.e., three
times the direction-averaged Hubble `constant' $H$), and then to do the integral
over $dH$.

Volume weighting has been the basis for the hypothesis that slow-roll eternal
inflation \cite{Vet,Let,Set,Guth00} makes the universe infinitely large (at
least with some nonzero probability \cite{CDNSZ}), that not only did the
universe undergo an early period of rapid quasi-exponential expansion, but also
that quantum fluctuations produce an unboundedly large amount of inflation by
today.  Therefore, abandoning volume weighting in favor of volume averaging
would apparently remove the argument for a nonzero probability for eternal
inflation to give infinite volume at the time of observers.  It would suggest
that although the universe may have inflated by a very large amount, it would
not have inflated by an infinite amount.  As a result, if space is compact, it
may well have a bounded volume with unit probability.  Such a finite universe
can more easily avoid the measure problem ambiguities of taking the ratio of
infinite quantities that occur in infinite universes arising from eternal
inflation.

\section{The measure problem in quantum cosmology}

A goal of quantum cosmology is to come up with one or more theories $T_i$ that
predict the probabilities of observations, by which I mean probabilities for the
results of observations.  Here for simplicity I shall assume that there is a
countable set of possible distinct observations $O_j$ out of some exhaustive set
of all such observations.  This set of possible observations might be all
possible conscious perceptions \cite{SQM,Page-in-Carr}, all possible data sets
for one observer, or all possible data sets for a human scientific information
gathering and utilizing system \cite{HS}.  (However, one should not mix
different types of observations that are not distinct from each other, such as
the data set for one person and the data set for all persons in some
civilization, since they are not distinct observations; one can give rise to the
other.)  If one imagines a continuum for the set of observations (which seems to
be logically possible, though not required), in that case I shall assume that
they are binned into a countable number of exclusive and exhaustive subsets that
each may be considered to form one distinct observation $O_j$.  Then the goal is
to calculate the probability $P(O_j|T_i)$ for the observation $O_j$, given the
theory $T_i$.

As noted in \cite{typder}, the mutually exclusive and exhaustive (complete) set
of possible observations and the probabilities assigned to them by the theories
should obey the following principles:

(1) {\it Prior Alternatives Principle\/} (PAP):

The set of alternatives to be assigned likelihoods by theories $T_i$ should
be chosen prior to (or independent of) the observation $O_j$ to be used to
test the theories.

(2) {\it Principle of Observational Discrimination\/} (POD):

Each possible complete observation should uniquely distinguish one element
from the set of alternatives.

(3) {\it Normalization Principle\/} (NP):

The sum of the likelihoods each theory assigns to all of the alternatives
in the chosen set should be unity,
\begin{equation}
\sum_j P(O_j|T_i) = 1  . 
\label{norm}
\end{equation}

One common strategy in cosmology is to calculate an unnormalized probability for
each possible observation within a fixed finite region and then try to form a
normalized sum of these over all regions of spacetime.  The problem arises when
there are an infinite number of regions and the sum of the unnormalized
probabilities for each region diverges when being added over all regions.  Then
it is not clear how to normalize the divergent sum to get unambiguous finite
ratios of the total probabilities for different possible observations.  A lot of
clever effort has gone into schemes for regularizing the total probabilities
[1-47]
but is seems fair to say that most of these schemes look rather {\it ad hoc} and
unnatural, at least to me.

One might think that once one has a quantum state for the universe, there would
be a standard quantum answer to the question of the probabilities for the
various possible observations.  For example \cite{typder}, one might take
standard quantum theory to give the probability $P(O_j|T_i)$ of the observation
as the expectation value, in the quantum state given by the theory $T_i$, of a
projection operator $\mathbf{P}_j$ onto the observational result $O_j$.  That
is, one might take
\begin{equation}
P(O_j|T_i) = \langle \mathbf{P}_j \rangle_i, 
\label{standard}
\end{equation}
where $\langle \rangle_i$ denotes the quantum expectation value of whatever
operator is inside the angular brackets in the quantum state $i$ given by the
theory $T_i$ (which I am taking not only to give the dynamics, e.g. the
Hamiltonian, or more generally the constraint equations of quantum gravity, but
also to give the boundary conditions, e.g., the quantum state itself in the
Heisenberg representation).  This standard approach works in the case of a
single laboratory setting where the projection operators onto different
observational results are orthogonal, $\mathbf{P}_j \mathbf{P}_k =
\delta_{jk}\mathbf{P}_j$ (no sum over repeated indices).

However \cite{typder,insuf}, in the case of a sufficiently large universe, one
may have observation $O_j$ occurring `here' and observation $O_k$ occurring
`there' in a compatible way, so that $\mathbf{P}_j$ and $\mathbf{P}_k$ are not
orthogonal.  Then the standard quantum probabilities given by Eq.
(\ref{standard}) will generically not obey the Normalization Principle Eq.
(\ref{norm}).  Thus one needs a different formula for normalizable probabilities
of a mutually exclusive and exhaustive set of possible observations, when
distinct observations within the complete set cannot be described by orthogonal
projection operators.

The simplest class of modifications of Eq. (\ref{standard}) would seem to be to
replace the projection operators $\mathbf{P}_j$ with some other `observation
operators' $\mathbf{Q}_j$ normalized so that $\sum_j \langle \mathbf{Q}_j
\rangle_i = 1$, giving
\begin{equation}
P(O_j|T_i) = \langle \mathbf{Q}_j \rangle_i . 
\label{eventual}
\end{equation}
Of course, one also wants $P(O_j|T_i) \geq 0$ for each $i$ and $j$, so one needs
to impose the requirement that the expectation value of each observation
operator $\mathbf{Q}_j$ in each theory $T_i$ is nonnegative.  If one wanted the
$\mathbf{Q}_j$'s to be independent of the theory $T_i$ and wished to allow
different $T_i$'s to give all possible quantum states, then it would be natural
to restrict the $\mathbf{Q}_j$'s to positive operators.  This is what I have
usually assumed, but it would be well to remember that it is just $\langle
\mathbf{Q}_j \rangle_i$ that needs to be nonnegative for each $i$ and $j$, and
that in principle the $\mathbf{Q}_j$'s themselves can depend on the theory
$T_i$.

The main point \cite{typder,insuf} is that in cosmology one cannot simply use
the expectation values of projection operators as the probabilities of
observations, so that each theory must assign a set of observation operators
$\mathbf{Q}_j$, corresponding to the set of possible observations $O_j$, whose
expectation values are used instead as the probabilities of the observations. 
Since these operators are not given directly by the formalism of standard
quantum theory, they must be added to that formalism by each particular complete
theory.  In other words, a complete theory $T_i$ cannot be given merely by the
dynamical equations and initial conditions (the quantum state), but it also
requires the set of observation operators $\mathbf{Q}_j$ whose expectation
values are the probabilities of the observations $O_j$ in the complete set of
possible observations.  The measure is not given purely by the quantum state but
has its own logical independence in a complete theory.

In quantum terms, the measure problem is the problem (say for giving a complete
formulation of a theory $T_i$) of choosing the observation operators
$\mathbf{Q}_j$ whose quantum expectation values are nonnegative and normalized
in the quantum state that is also to be given by the theory $T_i$.  The greatest
challenge in a theory giving an infinite universe (or a superposition of finite
universes for which the expectation value of the size is infinite) is the
normalization.

For example, assuming a classical spacetime as a suitable approximation, one
might hypothetically partition the spacetime into a countable set of disjoint
regions labeled by the index $K$, with each region sufficiently small that for
each $K$ separately there is a set of orthogonal projection operators
$\mathbf{P}_j^K$ whose expectation values give good approximations to the
probabilities that the observations $O_j$ occur within the region $K$. 
(However, I am assuming that each observation $O_j$ can in principle occur
within any of the regions, so that the content of the observation is not
sufficient to distinguish what $K$ is; the observation does not determine where
one is in the spacetime.  One might imagine that an observation determines as
much as it is possible to know about some local region of spacetime, or even
about the entire causal past of a local region, but it does not determine the
properties of the spacelike separated region outside, which might go into the
specification of the index $K$ that is only known to a hypothetical
superobserver that makes the partition.)

Now one might propose that one construct the projection operator
\begin{equation}
\mathbf{P}_j = \mathbf{I} - \prod_K (\mathbf{I} - \mathbf{P}_j^K)
\label{existence}
\end{equation}
and use it in Eq. (\ref{standard}) to get a putative probability of the
observation $O_j$ in the quantum state given by the theory $T_i$.  (For
simplicity I assume that the $\mathbf{P}_j^K$'s for different values of $K$ all
commute, so that $\mathbf{P}_j$ is indeed a projection operator.)  Indeed, this
is essentially in quantum language \cite{typder} what Hartle and Srednicki
\cite{HS} propose, that the probability of an observation is the probability
that it occur at least somewhere.  However, because the different
$\mathbf{P}_j$'s defined this way are not orthogonal, the resulting standard
quantum probabilities given by Eq. (\ref{standard}) will not be normalized to
obey Eq. (\ref{norm}).  This lack of normalization is a consequence of the fact
that even though it is assumed that two different observations $O_j$ and $O_k$
(with $j \neq k$) cannot both occur within the same region $K$, one can have
$O_j$ occurring within one region and $O_k$ occurring within another region. 
Therefore, the existence of the observation $O_j$ at least somewhere is not
incompatible with the existence of the distinct observation $O_k$ somewhere
else, so the sum of the existence probabilities is not constrained to be unity.

If one were the hypothetical superobserver who has access to what is going on in
all the regions, one could make up a mutually exclusive and exhaustive set of
joint observations occurring within all of the regions.  However, for us
observers who are confined to just part of the universe, the probabilities that
such a superobserver might deduce for the various combinations of joint
observations are inaccessible for us to test or to use to predict what we might
be expected to see in the future.  Instead, we would like probabilities for the
observations we ourselves can make.  I am assuming that each $O_j$ is an
observational result that in principle we could have, but that we do not have
access to knowing which region $K$ we are in.  (The only properties of $K$ that
we can know are its local properties that are known in the observation $O_j$
itself, but that is not sufficient to determine $K$, which might be determined
by properties of the spacetime beyond our local knowledge.)

A better proposal \cite{typder}, though only one out of a large number of
logically possible alternatives \cite{insuf}, would be to form the sum
\begin{equation}
\mathbf{R}_j = \sum_K \mathbf{P}_j^K
\label{R}
\end{equation}
of the projection operators $\mathbf{P}_j^K$ over all regions $K$ but for the
same observation $j$, and then to normalize this in the theory $T_i$ to
construct the positive observation operator
\begin{equation}
\mathbf{Q}_j = \frac{\mathbf{R}_j}{\sum_k \langle \mathbf{R}_k \rangle_i}
\label{Q}
\end{equation}
to be used in Eq. (\ref{eventual}) to give the normalized probability of the
observation $O_j$ in the theory $T_i$ as
\begin{equation}
P(O_j|T_i) = \langle \mathbf{Q}_j \rangle_i
= \frac{\langle \sum_K \mathbf{P}_j^K \rangle_i}
       {\sum_k \langle \sum_K \mathbf{P}_k^K \rangle_i}.
\label{EQMprob}
\end{equation}

This procedure appears to be the quantum language for what is usually done more
informally by adding up, over all regions, the unnormalized probabilities for
each region and then normalizing the result.  If one makes the approximation of
a uniform density of observations per volume, this procedure gives volume
weighting.  Then if one has a quantum state with different amplitudes for
different volumes (e.g., from stochastic inflation \cite{Vet,Let,Set,Guth00}),
the amplitudes for greater volumes (and hence a greater number of observers)
would be more greatly weighted.  Taking this to its logical conclusion
\cite{Vet,Let,Set,Guth00}, this leads to infinite volume in eternal inflation,
in which the probabilities for observations are dominated by the components of
the quantum state in which the universe has undergone an unbounded amount of
inflation.

Once one has a nonzero probability for the volume to diverge at the time of
observers or at reheating, as one does with volume weighting in eternal
inflation even for closed universes \cite{CDNSZ}, one has the problem of how to
produce well-defined normalized probabilities, say by finding well-defined
normalized observation operators $\mathbf{Q}_j$ if one uses quantum language. 
This is a manifestation of the measure problem.  Furthermore, there is the
danger that the result might be dominated by Boltzmann brains, making our actual
observations extremely unlikely in theories predicting such domination. 
Therefore, one is led to question the procedure of forming the observation
operators $\mathbf{Q}_j$ by summing over spacetime regions and then attempting
to normalize the result.

\section{Replacing volume weighting with volume averaging}

Volume weighting has been advocated by proponents of eternal inflation
\cite{Vet,Let,Set,Guth00}.  I have advocated it myself \cite{Page97} to try to
save the Hartle-Hawking no-boundary proposal from the fact that it gives the
highest bare amplitudes for universes with a small amount of inflation.  It has
been invoked in a similar way more recently by Hawking \cite{Haw} and by
Hartle, Hawking, and Hertog \cite{HHH}.  However, as we have seen, it gives the
problem of infinite total unnormalized probabilities and the ambiguities of how
to normalize the probabilities, as well as the danger of domination by Boltzmann
brains.

To help avoid these problems from domination by arbitrarily large volumes, I
propose replacing volume weighting by volume averaging.  That is, when
calculating the expectation value of $\mathbf{Q}_j$ over the entire quantum
state that is a superposition of different spatial geometries and matter fields,
weight the absolute square of the amplitude for that space and matter
configuration, not by the volume integral of the density of the occurrence of
the observation $O_j$, but rather by the volume-averaged value of the density. 
Then, for spatial geometries and matter field configurations that have the same
average density of the observation $O_j$, the weights contributing to
$P(O_j|T_i)$ are just the squared amplitudes in the quantum state for the
different spatial configurations, without any volume weighting.

This is the main new idea I would like to propose at present.  It is somewhat
{\it ad hoc} in that it cannot be derived from any basic underlying quantum
theory \cite{insuf}.  However, at the level of what measure to place on
observations occurring on a single hypersurface, volume averaging seems to be
the simplest alternative to volume weighting.  Therefore, if volume weighting
leads to trouble, volume averaging might appear to be an attractive alternative.

Unfortunately, we are not done yet with the problem.  To investigate the
consequences of volume averaging, one needs to say how one will add up the
measures from the various possible hypersurfaces.  For this part of the problem,
all the ideas I have come up with so far seem quite {\it ad hoc}, with many
conceivable alternatives and with many problems for each idea.  Therefore, I
certainly cannot claim to have a complete natural solution to the measure
problem.  However, in the hope that volume averaging might be part of an
ultimate solution when the problem of how to add the measures from different
hypersurface is solved, let me make some initial attempts to say how one might
do that addition.

In canonical quantum gravity, one would ultimately like to do an average over
all diffeomorphisms.  However, since the group of all diffeomorphisms is an
enormous infinite-dimensional set, I shall leave it as a future problem how to
implement this averaging over all diffeomorphisms to give the probabilities of
observations.  Here I shall merely seek an implementation for the approximation
that applies when the quantum state is considered to give an ensemble of
classical spacetimes with certain properties.  (However, there is no guarantee
that the quantum state will give an ensemble of classical spacetimes, just as
the quantum state of a hydrogen atom need not give an ensemble of classical
electron orbits, so ultimately we are likely to need a more quantum
implementation.)

\section{Hypersurfaces of constant mean extrinsic curvature $H$}

For each classical spacetime in an ensemble, rather than looking at all possible
hypersurfaces embedded in the spacetime, one might restrict attention to a
certain subset of these.  For example, one might look at a set of hypersurfaces
that foliate the spacetime, so that each spacetime event lies on one and only
only of these foliating hypersurfaces.

For globally hyperbolic spacetimes with compact Cauchy surfaces, York
\cite{York,MTW} has shown that the equations of general relativity simplify if
one uses a foliation by hypersurfaces of constant mean extrinsic curvature,
which one might call the direction-averaged Hubble `constant' $H$.  For
three-dimensional spatial hypersurfaces, this is one-third the trace of the
extrinsic curvature, which itself is the logarithmic expansion rate of the
volume of a comoving region generated by comoving worldlines orthogonal to the
sequence of hypersurfaces.  Here I shall call these hypersurfaces of constant
mean extrinsic curvature `constant-$H$ hypersurfaces.' Each of these
hypersurfaces has the advantage of being determined just by the spacetime
geometry in the neighborhood of the hypersurface (unlike, say, hypersurfaces
orthogonal to a congruence of timelike geodesics, which requires that the
congruence be specified throughout the spacetime and which also leads to
problems at caustics of the congruence).

A lot of effort \cite{York,York2,BriFla,MT,Ger,Bar,IR,Ak,Mon1,Ol,Mon2,Hen} has
gone into the analysis of the existence and uniqueness of a foliation by
constant-$H$ hypersurfaces, often under the assumption of the timelike
convergence condition \cite{HE} $R_{ab}u^au^b \geq 0$ for each timelike vector
$u^a$ at each point of spacetime.  This energy condition, which is equivalent to
the strong energy condition for matter when the Einstein equations hold with
zero cosmological constant \cite{HE}, was thought to be ``physically
reasonable'' \cite{HE} in the days before the popularity of inflation and the
observation of cosmic acceleration, both of which violate the timelike
convergence condition.

For example, York \cite{York} notes that in empty closed universes constant-$H$
hypersurfaces define ``a definite slicing of space-time.''  Marsden and Tipler
\cite{MT} prove their Theorem 3 that a nonflat spacetime having a smooth
spacelike compact Cauchy surface, containing a compact smooth spacelike maximal
hypersurface, and satisfying the timelike convergence condition, a hypersurface
generic condition, and development uniqueness, then has a unique foliation by
compact constant-$H$ hypersurfaces, with $H$ running from $+\infty$ at the big
bang to $-\infty$ at the big crunch, so long as constant-$H$ hypersurfaces avoid
singularities and don't turn null.  Gerhardt \cite{Ger} and Bartnik \cite{Bar}
have proved the existence of such a foliation under less stringent conditions
(though still assuming the timelike convergence condition), though Bartnik
\cite{Bar} also gives an example of a cosmological spacetime with compact Cauchy
hypersurfaces that obeys the timelike convergence condition and yet does not
have any constant-$H$ Cauchy hypersurfaces.  Later Isenberg and Rendall
\cite{IR} give an example that has a constant-$H$ Cauchy hypersurface but is not
covered by a foliation of such constant-$H$ hypersurfaces.

Less seems to be known about constant-$H$ hypersurfaces in spacetimes not
obeying the timelike convergence condition, such as inflationary spacetimes,
though the second paper cited by Brill and Flaherty \cite{BriFla} considers a
cosmological constant and discusses in detail the case of a Taub-type universe
that is homogeneous but anisotropic.  For de Sitter spacetime, Akutagawa
\cite{Ak} in three dimensions and Montiel \cite{Mon1} in higher dimensions
showed that compact constant-$H$ hypersurfaces are all umbilical, meaning that
all of the eigenvalues of the extrinsic curvature are equal.  These
hypersurfaces are all obtained by taking constant-time hypersurfaces in the
$k=+1$ FRW representation of de Sitter and performing de Sitter symmetry
transformations.  (On the other hand \cite{Ak,Mon1,Mon2}, there are complete
noncompact constant-$H$ hypersurfaces in de Sitter that are not umbilical and
hence not isometric to the umbilical noncompact hypersurfaces of constant time
in the $k=0$ and $k=-1$ FRW representations of de Sitter.)

The sequence of constant-time $k=+1$ hypersurfaces of de Sitter gives a
foliation of the entirety of de Sitter by compact constant-$H$ hypersurfaces,
but since one can perform de Sitter symmetry transformations of each of these
hypersurfaces, this foliation is not unique.  I might guess that a generic
small smooth perturbation of de Sitter would remove this ambiguity and lead to a
unique foliation by compact constant-$H$ hypersurfaces, but I am not aware of
any good evidence either for or against this guess.

For simplicity, here I shall initially assume that one restricts to ensembles
such that each spacetime within the ensemble may be foliated by complete compact
constant-$H$ Cauchy hypersurfaces.  Let $t$ be a time parameter that is constant
over each of these hypersurfaces and which increases monotonically as one goes
to the future along a sequence of these hypersurfaces.  Then one has $H = H(t)$,
a unique value of $H$ for each $t$, though if $H$ does not change monotonically
with $t$, one need not have a unique $t$ for each $H$ that occurs; there can be
more than one hypersurface with the same $H$.

Now to avoid dealing with the entire infinite-dimensional diffeomorphism group,
I would propose gauge fixing to this foliation.  One could also fix the time
coordinate $t$ (up to an overall shift by a constant) to be {\it volume-averaged
proper time} by letting $dt = dV_4/V_3$, where $V_3$ is the three-volume of the
constant-$H$ hypersurface at $t$ and $dV_4$ is the four-volume between that
hypersurface and the one at $t+dt$.  If there is a earliest infimum for $V_3$
(e.g., zero volume at a big bang, or a minimum volume at some smallest bounce
geometry), one could set $t=0$ there to eliminate the shift ambiguity in $t$.

Once one has fixed the foliation and the time parametrization, the remaining
diffeomorphism freedom is the coordinate freedom on the spatial hypersurfaces. 
In a neighborhood of any point on one of the hypersurfaces, one could always
gauge fix to Riemann normal coordinates, thereby reducing the
infinite-dimensional diffeomorphism group on that hypersurface to the
six-dimensional group generated by translations of the point and rotations of
the Riemann normal coordinates about that point.

Now suppose that instead of being a discrete index representing which region of
spacetime, $K$ is now taken to be an element of the six-dimensional group of
rotations and translations of the Riemann normal coordinates.  Then we might
imagine that $\mathbf{P}_j^K$ corresponds to a projection operator onto some
field configurations in the corresponding Riemann normal coordinate system that
would be or give rise to the observation $O_j$.  (For simplicity, I shall assume
that the region needed for the field configurations does not exceed the
applicability of the Riemann normal coordinates.)  Since the observation should
not depend on the orientation or location where it can occur (assuming that the
fields are located and oriented appropriately there), one might expect that the
contribution to $\mathbf{R}_j$ from one hypersurface (say $\mathbf{R}_j(t)$)
would be an average over the six-dimensional group of elements $K$ of the
$\mathbf{P}_j^K$ on the hypersurface at that value of $t$.  (The volume element
of the group can be taken to be the volume element of the three-dimensional
rotation group multiplied by the volume element of hypersurface for the
translations.  Since each hypersurface of the foliation is assumed to be
compact, the total volume of the group will also be compact, so the average over
the group should be well defined.)

In this way with gauge fixing to compact Cauchy hypersurfaces of constant
direction-averaged Hubble rate $H$, we can implement volume averaging over each
such hypersurface.  We still have to deal with the integral over different
spatial hypersurfaces.  It might be noted that by the preferred foliation of the
spacetime, we have avoided the noncompact integration over boosts in the Lorentz
group.  It might seem somewhat artificial to have avoided this, since one might
expect that in principle observations could be generated by configurations of
fields (including the fields of the observer) at any velocity relative to that
of the worldlines orthogonal to the preferred foliation by the constant-$H$
hypersurfaces.  However, I shall leave aside that issue for future analysis.

\section{Integrating over hypersurfaces}

In the approximation of regarding a quantum spacetime as an ensemble of
classical spacetimes, I have proposed gauge fixing to a particular foliation
(e.g., by hypersurfaces of constant trace of the extrinsic curvature, constant
logarithmic rate of the growth of three-volume) and then volume averaging on
those hypersurfaces.  It remains to say how to combine the effects of different
hypersurfaces in the foliation of each classical spacetime in the ensemble, that
is, how to integrate over the time parameter $t$ labeling the hypersurfaces in
the foliation.

The simplest proposal seems to be
\begin{equation}
\mathbf{R}_j = \int \mathbf{R}_j(t) dt,
\label{Rdt}
\end{equation}
where, as given above, $\mathbf{R}_j(t)$ is the group average of
$\mathbf{P}_j^K$ over the rotation group and volume of the hypersurface of fixed
$t$ and constant $H(t)$, and where $dt$ is the volume-averaged proper time
between nearby hypersurfaces.  This proposal might be called proper-time
weighting of the hypersurfaces (in distinction to the volume averaging of
observations made within each hypersurface that is my main proposal, though both
parts are needed for a complete specification of the measure).

Another proposal would be
\begin{equation}
\mathbf{R}_j = \int \mathbf{R}_j(t) |dH/dt| dt,
\label{RdH}
\end{equation}
the sum of $\int \mathbf{R}_j(t) |dH|$ over all segments of the history where
$H$ changes monotonically with the volume-averaged proper time $t$.  This
proposal might be called Hubble-constant-time weighting of the hypersurfaces.

A third proposal is
\begin{equation}
\mathbf{R}_j = \int \mathbf{R}_j(t) dV_4 = \int \mathbf{R}_j(t) V_3(t) dt,
\label{RdV}
\end{equation}
where $dV_4$ is the four-volume element between infinitesimally nearby
hypersurfaces in the foliation, and $V_3(t)$ is the three-volume of the
hypersurface in the foliation at $t$.  This proposal might be called four-volume
weighting of the hypersurfaces.  It essentially restores the three-volume
weighting.  Because of the problems I have noted with volume weighting, here I
shall not advocate this four-volume weighting procedure of the hypersurfaces,
but I am pointing it out to show that it is not {\it a priori} obvious to me
that one weighting is intrinsically far superior to another.  This fact may just
be a manifestation of what I have noted above \cite{insuf}, that for predicting
probabilities of observations, a theory is not fully specified by just the
dynamical laws and boundary conditions, but rather one also needs the detailed
procedure for calculating the probabilities of observations.  Therefore, it may
be a mistake to expect this procedure to arise intrinsically from the formalism
for the dynamical laws and boundary conditions.  One needs more than just the
quantum dynamics and the quantum state.

\section{Problems with Constant-$H$ Hypersurfaces}

Although compact constant-$H$ hypersurfaces seem to be the simplest prescription
for foliating a spacetime when they work, there are problems with them.  First
of course, they certainly will not work to foliate a spacetime that is not
globally hyperbolic or that does not have compact Cauchy hypersurfaces. 
However, I have no good idea how to handle such cases at all, so for now I shall
assume that such spacetimes can be excluded.  (Perhaps they never arise from
approximations to the quantum state of the universe, which conceivably could
give only globally hyperbolic spacetimes with compact Cauchy hypersurfaces, if
it gives spacetimes at all in some level of approximation.)

Second, there are known examples \cite{Bar,IR} mentioned above in which globally
hyperbolic spacetimes with compact Cauchy hypersurfaces cannot be foliated by
compact constant-$H$ hypersurfaces.  One might regard these examples as also
rather pathological, but there is also a common example in inflation that
apparently cannot be completely foliated by compact constant-$H$ hypersurfaces: 
de Sitter with the formation of a thin-wall Coleman-De Luccia \cite{CDL} bubble
of Minkowski spacetime inside.  I-Sheng Yang \cite{Yang} has shown that a
foliation by compact constant-$H$ hypersurfaces covers only a small part of the
Minkowski spacetime region inside the bubble and also avoids an infinite
spacetime volume in the de Sitter region outside the bubble (though it also
covers an infinite spacetime volume in the de Sitter region, up to infinite
proper time along a large set of timelike worldlines that stay outside the
bubble).  Indications suggest that the similar behavior may occur if a
Coleman-De Luccia bubble of our value of the cosmological constant formed out of
a parent de Sitter universe of a larger cosmological constant.  Then the compact
constant-$H$ hypersurfaces that cross the parent de Sitter region would not
enter into our part of the spacetime nearly far enough to cover our existence.

If this behavior is indeed confirmed by a more careful analysis, and if indeed
our part of spacetime formed as a bubble from some pre-existing spacetime with a
much larger cosmological constant, it would suggest that our existence would not
contribute to the measure defined in terms of a foliation by compact
constant-$H$ hypersurfaces.  However, I am a bit sceptical of the usual picture
in eternal inflation that our part of spacetime most probably formed from
tunneling from some previous part of spacetime with a larger cosmological
constant.  I would prefer to explore the alternative that our part of spacetime
formed rather directly, without having had a significant probability to have had
ancestor regions of spacetime with different cosmological constants.  (Although
I am a creationist in the sense of believing the universe was created by God, I
am not a recent-creationist in the traditional sense of believing that the
universe was created within the past ten thousand years or so.  But am I a
recent-creationist in the sense that I suspect that the universe may most
probably have started around 14 billion years ago instead of far earlier before
some long series of tunnelings from some original ancestor spacetime to our
pocket universe?  My main worry with the latter popular scenario is not
theological but whether the arrow of time would persist through an arbitrarily
long sequence of bubble formations and decays, or whether the entire multiverse
would tend toward some sort of heat death that would be inconsistent with our
observations.)

If the main contribution to the path integral for our present existence comes
from our own spacetime region or pocket universe, without the need for a
sequence of tunnelings from some ancestor region, then our existence might
indeed be covered by some foliation of our region by compact constant-$H$
hypersurfaces.  If our pocket universe decays into future bubbles of smaller or
negative cosmological constant (e.g., terminal vacua), then the constant-$H$
foliation may indeed not penetrate very far into those regions, and it might not
cover all of the future of our apparently asymptotically de Sitter region, but
if it covers the existence of most observers in our part of spacetime, that
should be sufficient to get at least approximately the right measure for
our observations.

\section{Alternatives to Constant-$H$ Hypersurfaces}

Although constant-$H$ hypersurfaces presently seem to be the simplest known and
most studied proposal for foliating at least some large class of spacetimes, I
would certainly not claim that they are definitely the correct way to do
things.  If indeed they do not work (which does now seem rather likely if our
pocket universe really did come from a long sequence of decays of previous
vacua), one should look for alternatives.

A common alternative is to postulate some initial smooth spacelike Cauchy
surface, construct timelike geodesics orthogonal to it, and then define
foliating hypersurfaces to be the orthogonal hypersurfaces to this congruence of
geodesics.  One disadvantage of this procedure is the requirement of the choice
of the initial hypersurface (or alternatively simply of the congruence itself),
but this might be regarded as a small price to pay if the result overcomes any
severe difficulties that might arise from other proposed foliations.

A much more severe problem with this proposed foliation is that generically the
timelike geodesics will have conjugate points where they will begin to cross and
form caustics.  This will prevent the orthogonal hypersurfaces from remaining
smooth and from being crossed only once by each inextendible timelike geodesic. 
Furthermore, there will be more than one hypersurface at many points of
spacetime, where different timelike geodesics from the initial hypersurface
intersect with different values of proper time.  Within the solar system,
timelike geodesics that remain within the sun or a planet will typically have
conjugate points separated by only an hour or so, making the hypersurfaces
orthogonal to the geodesics run into difficulty very quickly.  Therefore, some
modification of the procedure will be needed.

One modification that can partially ameliorate the problem is not to require
that the hypersurfaces be orthogonal to the entire congruence of crossing
geodesics, but instead to have constant values of a time function that is
defined to be the maximum proper time of any causal curve back to the original
hypersurface.  The curves that maximize the proper time will of course be
timelike geodesics that intersect the original hypersurface orthogonally, but
for later points intersected by more than one geodesic orthogonal to the
original hypersurface, only the longest one will count for defining the time
function.  Thus one will get definite spacelike hypersurfaces of constant proper
time from this procedure.

One disadvantage of this procedure for getting constant-proper-time
hypersurfaces is that generically on the three-dimensional hypersurfaces of
constant maximal proper time after the original hypersurface, there will be
two-dimensional surfaces where two such maximal geodesics intersect.  On one
side of the two-surface, the time function will be given by one local
congruence, but on the other, it will be given by a different local congruence. 
The two congruences will have different tangent vectors (four-velocities) on the
two sides, so at the two-dimensional surface, the normal vector to the
three-dimensional hypersurface will jump by some boost, making the constant-time
hypersurface have a kink.  There can also be further defects along
one-dimensional lines within the hypersurface, as well as at isolated points. 
But if one can live with a foliation by hypersurfaces with kinks and other
defects, and if one is willing to specify an original hypersurface, then the
ones of constant proper time from the original hypersurface would seem to work
in any globally hyperbolic spacetime with compact Cauchy hypersurfaces.

Another alternative worth exploring is what might be called `minimal-flux
hypersurfaces.'  For a given smooth compact Cauchy hypersurface with unit
future-pointing timelike normal $n^a$, the energy density in the frame of an
observer with 4-velocity $n^a$ is $\rho = T_{ab}n^an^b$, and the spatial energy
flux in the frame of this observer is $J^a = -T^a_{\ b}n^b - \rho n^a$
\cite{Wald}.  First, consider only hypersurfaces for which the volume average of
the energy density $\rho$ (using the 3-volume element from the metric induced on
the hypersurface by the spacetime metric) has a fixed value $\bar{\rho}$.  Next,
for such hypersurfaces, choose the one that minimizes the volume average of the
square of the spatial energy flux, $J^aJ_a$.  Assuming the existence and
uniqueness of such a hypersurface, it can be defined to be the minimal-flux
hypersurface for the given value of the average energy density $\bar{\rho}$. 
Finally, consider the one-parameter sequence of such compact minimal-flux
hypersurfaces as a function of $\bar{\rho}$.  If this indeed foliates the
spacetime, this will be a `minimal-flux foliation.'

In the case that the stress-energy tensor $T_{ab}$ has a unit timelike
eigenvector field $u^a$ (obeying $u^au_a = -1$ and $T_{ab}u^a = -\hat{\rho}u_b$
for some eigenvalue $\hat{\rho}$, which is the energy density in the frame with
4-velocity given by the eigenvector $u^a$, a frame in which there is no spatial
energy flux), and in the case that this eigenvector field is
hypersurface-orthogonal, then one may choose the hypersurfaces so that $n^a =
u^a$, giving $\rho = \hat{\rho}$ and $J^a = 0$, so such hypersurfaces would be
minimal-flux hypersurfaces.  (For example, in $k=+1$ FRW spacetimes, the
homogeneous isotropic hypersurfaces of constant $t$ would be such compact
minimal-flux hypersurfaces.)  However, generically the timelike eigenvector
field (if it exists, which is also not guaranteed but seems to be the generic
case) would not be hypersurface orthogonal, so that there would not be
hypersurfaces with zero spatial energy flux $J^a$ everywhere.  Minimal-flux
hypersurfaces make one particular minimization of the deviation of their normals
$n^a$ from the timelike eigenvector field $u^a$.

It would be worth exploring more the properties of minimal-flux hypersurfaces
and the cases in which they do foliate part or all of a spacetime with compact
Cauchy hypersurfaces.  For example, do they foliate generic perturbations of FRW
spacetimes?  If a classical approximation to our quasiclassical component of the
quantum state of the universe does have a minimal-flux hypersurface through our
present location in this spacetime, what is the deviation of $n^a$ from $u^a$ at
the surface of the earth?  (I might guess that would be of the order of the
magnitude of the peculiar velocity of our galaxy from the mean Hubble flow, but
one might ask whether the part of spacetime far beyond what we can see could
possibly force the minimal-flux hypersurface to have a significantly greater
variation of $n^a$ from $u^a$ at our location.)

Minimal-flux hypersurfaces would seem to have the advantage that their normals
$n^a$ would presumably be close to the 4-velocity $u^a$ of the local matter
frame (say given by the unit eigenvectors of the stress-energy tensor), at least
for situations in which the peculiar velocities of the matter are small, as is
generally the case in our part of the universe.  A foliation by these
hypersurfaces thus might avoid the objection Vilenkin \cite{Vilenkin} has raised
to the constant-$H$ hypersurfaces, that inside new bubble universes their
normals would typically be at high velocities to that of the matter there, even
if the bubbles were locally close to open FRW universes.  (As noted above, it
now looks \cite{Yang} as if the problem is even more severe for such bubbles, in
that apparently the compact constant-$H$ hypersurfaces do not even go very far
into the new bubble universes.)  Although I am not convinced that the dominant
contribution to the measure for observations will come from baby universes
arising from bubble nucleation out of a parent universe (as opposed to direct
creation of our pocket universe without any ancestors), if one does want to
investigate the measure inside baby bubble universes, minimal-flux hypersurfaces
might be a better attempt for a foliation.  However, this has not yet been
investigated.

One difficulty in analyzing minimal-flux hypersurfaces is that they rely for
their definition on the stress-energy tensor and so are not defined for vacuum
spacetimes, but that should not be a problem for any spacetime region rather
like our own, which certainly does have matter and a nonzero stress-energy
tensor.  There might conceivably be difficulties in an asymptotically de Sitter
future part of spacetime with a cosmological constant as the non-vacuum part of
the stress-energy tensor tends toward zero.  (A purely vacuum part proportional
to the metric, such as the cosmological constant, gives $J^a = 0$ for all
hypersurfaces and therefore provides nothing nontrivial to minimize).  However,
one might imagine that minimal-flux hypersurfaces would be defined wherever
there is any reasonable nonzero stress-energy tensor, no matter how small.  A
bigger worry in my mind is whether the sequence of them with smaller and smaller
$\bar{\rho}$ continue to foliate the spacetime, or whether these hypersurfaces
might cross each other or else simply not cover all of the spacetime.

Clearly constant-$H$ hypersurfaces, constant-proper-time hypersurfaces, and
minimal-flux hypersurfaces are all rather {\it ad hoc} proposals.  Although I am
happy to propose volume averaging rather than volume weighting for observations
within a hypersurface, I am not happy with anything I have thought of so far for
what hypersurfaces to include and for how to add up their contributions.  This
is definitely an ugly part that needs some elegant new ideas.

\section{Violation of the Correspondence Principle}

One objection \cite{Vilenkin} to volume averaging rather than volume weighting
is that it violates the ``correspondence principle'' that for a finite
spacetime, ``The probability of a given outcome of a measurement can be simply
defined as the relative number of instances when this outcome is obtained.'' 
For example, consider a $k=+1$ FRW spacetime with both a big bang and a big
crunch, so that the scale factor $a(t)$ goes to zero at both $t=0$ and $t=T$,
where $t$ is proper time of the longest timelike curve from the big bang, and
$T$ is the finite lifetime of this universe.  In this simple case the
constant-$H$ hypersurfaces will be the usual homogeneous, isotropic $S^3$
hypersurfaces of fixed proper time $t$ and 3-volume $V_3 = 2\pi^2 a^3(t)$, and
the volume-averaged proper time will be this same $t$, since $dV_4 = V_3 dt$. 
Therefore, let us use the foliation by constant-$H$ hypersurfaces and then
compare four-volume weighting with proper-time weighting of the hypersurfaces.

Now let $n(t)$ be the density per 4-volume of some observation, for simplicity
uniform over each constant-$H$ hypersurface.  Then $N_{VW} = \int_0^T 2\pi^2
a^3(t) n(t) dt$ would be the total number of these observations in the
spacetime, which is what one would get by volume weighting (equivalent to volume
averaging over each hypersurface and then four-volume weighting for the integral
over the hypersurfaces).  On the other hand, volume averaging over each
hypersurface and then proper-time weighting of the hypersurfaces would give a
total weight for the observation proportional to $N_{VA} = \int_0^T n(t) dt$,
simply the proper-time integral of the average density of observations.  Thus
volume averaging gives less weight to observations when the hypersurfaces have
greater volume, violating the ``equivalence principle.''

I readily admit that this is indeed a feature of volume averaging.  However, it
is not clear that it is in conflict with any observational or theoretical
requirement.  We have no strong evidence that the probability measures for
individual observations are not actually going down as the volume of space goes
up.  However, if we use hypersurfaces of constant proper time since decoupling
(or since reheating, assuming inflation occurred earlier) in our part of the
universe, the volume has gone up by only a tiny fraction during the history of
the entire human race, by a factor less than 1.00005 during the $200\,000$ years
of modern humans.  Perhaps, for other things being equal, our present
observations have 0.005\% lower measure than those of early modern humans, but I
don't see how we could possibly exclude this possibility observationally.

There would be a serious problem if our pocket universe had developed as a
bubble in a parent universe with a much greater cosmological constant and if the
hypersurfaces on which one does the volume averaging have the main contribution
to their volume from the rapidly expanding parent universe region.  Then the
total volume of the hypersurfaces through us (most of the volume being in the
rapidly expanding parent universe) would be growing at a rapid logarithmic rate,
so with volume averaging the weight for our observations would rapidly decrease
with time.  This would then lead to the youngness paradox
\cite{LLM,Guth00,Teg,BFYa}.  However, as discussed above, I prefer to explore
the alternative that our pocket universe formed rather directly, most probably
without ancestors, so that the main contribution to the volume of the
hypersurfaces lies within our pocket universe itself and is not growing very
rapidly on human scales.  Then the violation of the ``correspondence
principle,'' although in principle present, would be very tiny during the past
lifetime of human civilization.

\section{Application to Boltzmann brains}

Replacing volume weighting by volume averaging can greatly ameliorate the
problem of Boltzmann brains, though whether it completely solves the problem
depends on the procedure for the integration over hypersurfaces and on whether
the universe lasts forever.  I shall here leave aside the four-volume weighting
procedure, since that seems to leave the problem in the severe form it has with
the usual volume weighting.

Suppose that the probability per Planck volume of a Boltzmann brain is very
roughly $e^{-I}$, where $I$ is much, much larger than the value of the
logarithm, say $J$, of the reciprocal of the probability per Planck volume of an
ordinary observer in the present universe.  For example, $I$ might be the
instanton action for producing a Boltzmann brain.  If one wants a brain of mass
and size comparable to that of a human brain, say mass of the order of 1 kg and
size of the order of 0.3 m, then $I \sim 10^{42}$ for a `brief brain'
\cite{Page06b}.  If one wants a brain of mass 1 kg to last 0.1 second, which is
what some people think may be necessary for it to make an observation, then $I
\sim 10^{50}$ for a `medium brain' \cite{Page05,Page06b}.  On the other hand, if
one wants a brain of mass 1 kg to last for several Hubble times (or to be
produced as a real thermal fluctuation in the future asymptotic de Sitter
spacetime), then $I \sim 10^{69}$ for a `long brain' \cite{Page06b}.

Then in the universe up to the present time, the fraction of Boltzmann brains to
ordinary observers is very roughly $e^{-I+J}$, so we can ignore their effect
from just the past part of our spacetime.  But now suppose ordinary observers
mostly die out when the stars burn out and do not last enormously longer,
whereas Boltzmann brains continue to fluctuate into and out of existence at
their very tiny rates.

If we had used volume weighting, Boltzmann brains would dominate when the
spatial volume is larger than today by a factor of roughly $e^{I-J}$.  Since the
volume asymptotically grows at a rate $e^{3H_\lambda t}$, with $H_\lambda =
\sqrt{\Lambda/3}$ in terms of the cosmological constant $\Lambda$, this time is
of the order of $t \sim (I-J)/(3H_\lambda)$, of the order of $10^{52}$ years if
`brief brains' can have observations (as I myself would think is most
plausible), $10^{60}$ years if one needs `medium brains,' or $10^{79}$ years if
one needs `long brains' \cite{Page05,Page06b}.  Then if the universe is
destroyed or decays deterministically before the appropriate one of those times,
there would not be a Boltzmann brain problem.  However, it is hard to see what
is likely to make the universe cease to exist at such a time that is so short in
comparison with the Poincar\'{e} recurrence time of the order of $e^{10^{123}}$
years for de Sitter spacetime with the presently observed value of the
cosmological constant.  One gets an even much shorter expected decay time of the
order of $10^{10}$ years to prevent Boltzmann brains from dominating with volume
weighting if the universe decays quantum mechanically by bubble formation rather
than deterministically \cite{Page06a,Page06d}, since with a lower decay rate,
the expectation value of the four-volume diverges, giving an expectation value
of Boltzmann brains per comoving volume that is infinitely more than of ordinary
observers in our universe.  (This result may be modified by a consideration of
ordinary observers and/or Boltzmann brains that may form in the next universe
after the bubble decay \cite{Linde06,Page06c}, but the answer is not clear
because of the ambiguities of the measure problem with volume weighting.)

If we use instead proper-time weighting of hypersurfaces without volume
weighting for observations on hypersurfaces, we need the expectation value of
the total proper-time lifetime of the universe (rather than of the four-volume)
to be less than roughly $t_0 e^{I-J}$.  If we just try to get the order of
magnitude of the exponent roughly right, this gives $e^{10^{42}}$ years for
`brief brains,' $e^{10^{50}}$ years for 'medium brains,' or $e^{10^{69}}$ years
for `long brains.'  Although these times are also enormously shorter than the
Poincar\'{e} recurrence time for de Sitter spacetime, the logarithms of their
logarithms are of the same order of magnitude, so it might be plausible that
spacetime would decay within one of those timescales \cite{CDGKL,FL}.  But if
the universe does last forever in a form wherein Boltzmann brains can continue
to fluctuate into (and out of) existence, then it appears that there is still a
problem with Boltzmann brains even without volume weighting if one uses
proper-time weighting.

Even if the universe does have an expected decay time shorter than say
$e^{10^{42}}$ years, there still may be a problem \cite{Bousso} if the decay is
by bubble formation of terminal vacua that leaves the bulk of our asymptotically
de Sitter spacetime intact.  Then presumably the constant-$H$ (or constant
proper time) hypersurfaces will continue to evolve forward in the expanding part
of de Sitter spacetime that remains outside the bubble formation, so that there
would be an infinite time for Boltzmann brains to appear on it.  One might think
\cite{Yang} that the volume fraction of the hypersurfaces that stay outside the
bubbles would decrease fast enough to lead to a convergence in the integral over
$dt$ of the average density of Boltzmann brains on the hypersurface (presumably
only on the part that stays outside the terminal bubbles), but if the
three-volume inside each bubble is small, whereas the part of the hypersurface
outside keeps expanding with the asymptotically de Sitter universe, it seems to
me more plausible that the fraction of the three-volume inside bubbles would
always remain small.  Then the density of Boltzmann brains, averaged over the
entire hypersurface, would remain near its nearly constant value in the
asymptotically de Sitter part, and so the integral of that over $dt$ would
diverge along with $t$.

I would like to postulate that when terminal vacua are forming, somehow the
quantum measure for the hypersurfaces decreases exponentially.  If one imagined
some sort of collapse of the wavefunction to possible nonexistence over the
entire hypersurface with some probability proportional to the probability that a
bubble forms on the hypersurface, this might be accomplished, but I am highly
sceptical of any collapse of the wavefunction that would occur acausally over a
hypersurface.  I suppose one could still postulate a decrease in the absolute
value of the existence amplitude of the hypersurface, as if it represented a
probability for it to collapse out of existence, without invoking any actual
collapse.  However, for this to be caused by the amplitude for a bubble to occur
at some location on the hypersurface smacks of some degree of acausality.  On
the other hand, a quantum state is a function or functional of an entire
configuration space and so is certainly nonlocal.  If the quantum state changes
according to what may be happening on the hypersurface (e.g., by the potential
for bubble formation), then since the quantum state is inherently nonlocal, it
may appear as if the change is acausal.

If it turns out that the proper-time weighting of hypersurfaces simply does not
work, one might instead turn to the Hubble-constant-time weighting of
hypersurfaces.  Then the integral at late times will not diverge at all, since
$H(t)$ is expected to drop from its present finite value to a finite asymptotic
value $H_\Lambda = \sqrt{\Lambda/3}$ if the dark energy driving the current
cosmological acceleration is a cosmological constant or a minimum in a scalar
field potential, or else to the finite asymptotic value of zero if the dark
energy decays away.  One might expect a divergence instead at infinite values of
$H(t)$ if the universe started at a big bang with an infinite logarithmic
expansion rate.  However, if $H(t)$ has an upper cutoff at the Planck value,
that would also keep the integral convergent and still sufficiently small when
weighted by $e^{-I}$ that Boltzmann brains would not come at all close to
dominating.

Even if one took $H(t)$ back to infinity at a classical big bang at $t=0$, one
might argue that in a region large enough for a Boltzmann brain, the energy
density would be so high that the probability would be exponentially small that
it would take the form of a Boltzmann brain.  For example, suppose the entropy
goes as a positive power $p$ of the energy density, which itself is expected to
go as $t^{-2}$ for small $t$ near the big bang, so that the entropy in a region
of the volume of a Boltzmann brain goes as $S(t) \sim Ct^{-2p}$.  Then if there
are only a finite number of configurations in a given volume that would
correspond to Boltzmann brains, then the probability that one of the $e^S$
configurations would be a Boltzmann brain would go roughly as $P(t) \sim
e^{-S(t)} \sim e^{-Ct^{-2p}}$.  Now if the scale size goes as some positive
power $q$ of the proper time, $a(t) \sim At^q$, so that $H(t) = \dot{a}/a \sim
q/t$ and $|dH/dt| \sim q/t^2$, then $\int P(t) |dH/dt| dt \sim \int
e^{-Ct^{-2p}} q t^{-2} dt$ clearly does not diverge at $t=0$.  Almost certainly
with the $|dH/dt|$ weighting factor, the probability of a observation by a
Boltzmann brain in a big-bang universe with Hubble-constant weighting would be
far below that of an ordinary observer.  Thus the Boltzmann brain problem would
be solved with volume averaging of observations on each hypersurface and with
Hubble-constant-time weighting of hypersurfaces, even if the universe lasts
forever.

On the other hand, in \cite{HP} Hawking and I argued that ``In situations in
which the wave function can be interpreted in terms of classical solutions by
the WKB approximation, this choice of measure implies that the probability of
finding the 3-metric and matter field configuration in a given region of
superspace is proportional to the proper time that the solutions spend in that
region."  This argument would tend to support the proper-time weighting of
hypersurfaces.

\section{Application to Proposed Quantum States of the Cosmos}

Volume averaging (rather than volume weighting) for observations on
hypersurfaces thus seems to help solve the measure problem and the Boltzmann
brain problem (at least if the universe does not last extraordinarily long, or
if one uses Hubble-constant-time weighting of hypersurfaces).  However, that is
only the case if the quantum state of the universe is dominated by finite-volume
spaces that evolve from a big bang (or at least from a region of high density)
and then expand to give ordinary observers with high probability.  There can
still be problems explaining our observations if the quantum state is dominated
by spaces without ordinary observers, such as large nearly empty spaces.

For example, the Hartle-Hawking `no-boundary proposal' still seems problematic
\cite{Page06b}, because it gives an enormous amplitude for nearly empty de
Sitter spaces, whose Boltzmann brains would greatly dominate over the ordinary
observers that only exist in a part of the quantum state with much lower
amplitude, even without volume weighting.  The `tunneling' proposals of
Vilenkin, Linde, and others
[76, 103-108]
seem to fare better, though the ones of these that just reverse the sign of the
Euclidean action give divergent amplitudes for arbitrarily large perturbations 
[109-111]
(which is not what Vilenkin's tunneling proposal does \cite{V,Vet,TVV,TGV},
though one might say that this proposal is not yet precisely defined, even at
the level of minisuperspace).

The no-bang proposal \cite{nb} for the quantum state of the universe appears to
avoid some of the problems of the no-boundary proposal and yet seems to be more
precisely defined than the tunneling proposal.  However, without volume
weighting, the no-bang state appears to be dominated by thermal perturbations
of nearly empty de Sitter spacetime, in which almost all observers would
presumably be Boltzmann brains.  Since this would almost certainly make our
observations very unlikely, the no-bang proposal apparently is observationally
excluded if one uses volume averaging rather than volume weighting.

Therefore, for volume averaging to solve the measure problem and avoid the
Boltzmann brain catastrophe, one needs a quantum state that is not enormously
dominated by nearly empty de Sitter spacetime.  One would like a state that is
dominated by spacetimes having a big bang or bounce at volumes much smaller
than that given by the apparent cosmological constant observed today.  Clearly
more work needs to be done to find such a state.

\section{Conclusions}

The measure problem is a severe problem in theoretical cosmology that is
logically independent of the question of what the quantum state of the universe
is.  With a suitable quantum state, replacing volume weighting with volume
averaging in the cosmological measure appears to help avoid the ambiguities of
infinite measures generated by eternal inflation and also avoid the Boltzmann
brain catastrophe and the youngness paradox.  It also appears to avoid the ``$Q$
catastrophe'' of exponentially preferring either huge or tiny primordial density
contrasts \cite{FHW,GV06} and the analogous catastrophe for the gravitational
constant $G$ \cite{GS}.  However, even with this volume averaging measure, the
result does depend on the quantum state, and one does still need to find a
quantum state that would give sufficiently high probabilities for our
observations.

One consequence of volume averaging rather than volume weighting would be a loss
of the argument for infinite volumes today from eternal inflation
\cite{Vet,Let,Set,Guth00,CDNSZ}.  There would still be amplitudes for
arbitrarily large amounts of inflation, but without volume weighting, the bulk
of the probabilities for observations would occur for spaces with only a bounded
amount of inflation.  However, although there is a lot of indirect observational
evidence for inflation itself, I think it is fair to say that so far there is
not any observational evidence for infinite volumes from eternal inflation in
particular.  (It is hard to see how we can have any direct observational
evidence about the volume of the spatial hypersurface containing us, since all
of it outside ourselves is acausally related to us now.)  It would be
interesting to see whether there is any observational way to confirm or refute
infinite volumes from eternal inflation, other than the Boltzmann brain
catastrophe that often seems to accompany theories of infinite volume from
eternal inflation.  Volume averaging can help kill both Boltzmann brains and
infinite volumes from eternal inflation, but it remains to be see whether this
solution can be observationally distinguished from solutions that kill Boltzmann
brains but not infinite volumes from eternal inflation.

Many of the implications of volume averaging rather than volume weighting are
qualitatively similar to those of the causal diamond or holographic point of
view of Bousso and collaborators \cite{Bou06,BFY06,BF,BHKP,BY,BFYa}.  They argue
that one should restrict attention to the causal diamond region that is both in
the past and in the future of an observer's history or worldline.  I would agree
that perhaps the only testable predictions of a theory involve such restricted
regions.  (I would be even more radical in proposing that the predictions should
involve only the data that has entered the observer and no external data at
all).  However, to me it would seem that one should allow a theory making such
predictions to give a more global description of the quantum state of the
universe.  (This is not a claim that there should be some classical global
spacetime for the entire universe.  Surely there will be a huge quantum
superposition.)  If volume averaging can avoid the problems of volume weighting
that the causal diamond approach also avoids, and yet keep a global description,
then it would seem advantageous

It appears that yet another way to get rather similar results is the
scale-factor cutoff measure \cite{SGSV,BFYb,DGLNSV}.  It is not yet clear which
of these various approaches is best.  However, I would suspect that as an
approximation to a prescription best given in a quantum analysis, it would be
better to try to avoid properties of a global classical spacetime, such as
any time parameters defined by some classical history of spacetime.  For this
reason I am not convinced that global time cutoffs are the best approach.  This
is my motivation for looking at hypersurfaces instead and trying to define them
quasi-locally, since three-metrics and matter fields on spacelike hypersurfaces
are the traditional configuration-space coordinates in canonical quantum
gravity.  (Of course, there are severe problems with canonical quantum gravity,
including the fact that with fluctuations in the geometry it may not be definite
that any hypersurface is acausal, allowing quantum xeroxing to occur and
preventing a description of the quantum state as a wavefunctional of
hypersurface geometries and matter fields.)

In any case, it may be premature to say which, if any, of the current approaches
is most likely to lead to the ultimate answer to the measure problem.  At
present it seems best to investigate a variety of approaches, see what their
consequences are, and try to find whether one can find a truly quantum
implementation.  It seems that volume averaging of observations on
hypersurfaces, though by no means complete without a specification of how to do
the sum over hypersurfaces, merits further investigation.

\section*{Acknowledgments}

I am grateful for discussions with Andreas Albrecht, Anthony Aguirre, Tom Banks,
Nick Bostrom, Raphael Bousso, Steve Carlip, Bernard Carr, Sean Carroll, David
Coule, William Lane Craig, George Ellis, Ben Freivogel, Claus Gerhardt, Gary
Gibbons, Steve Giddings, J. Richard Gott, Alan Guth, Jim Hartle, Stephen
Hawking, Thomas Hertog, Jim Isenberg, Renata Kallosh, Pavel Krtou\v{s}, John
Leslie, Andrei Linde, Juan Maldacena, Robert Mann, Don Marolf, Jens Niemeyer,
Joe Polchinski, Martin Rees, Michael Salem, Mark Srednicki, Glenn Starkman,
Leonard Susskind, Max Tegmark, Neil Turok, Bill Unruh, Alex Vilenkin, I-Sheng
Yang, and James W. York, Jr.  Email debates with Hartle and Srednicki have
helped me, at least, come to the conclusion that the measure problem is
logically separate from the problem of what the quantum state of the universe is
\cite{insuf}.  I especially appreciate a detailed critique of the previous
version of this paper by Alex Vilenkin, and the hospitality of the University of
California at Berkeley to discuss this work with Raphael Bousso, Ben Frievogel,
Jens Niemeyer, and I-Sheng Yang.  This research was supported in part by the
Natural Sciences and Engineering Research Council of Canada.

\end{document}